\documentclass[11pt,a4paper]{article}
\usepackage{charter,url}
\usepackage[utf8]{inputenc}
\usepackage[bottom=3cm,top=3cm,left=3cm,right=3cm]{geometry}
\usepackage{graphicx}
\usepackage{amsmath}
\usepackage{mathrsfs}
\usepackage{amsfonts}
\usepackage{amssymb}
\usepackage{bbm}
\usepackage{endnotes}
\usepackage{hyperref}
\usepackage{color}

\newcommand{\be}{\begin{equation}}
\newcommand{\ee}{\end{equation}}
\newcommand{\ben}{\begin{equation*}}
\newcommand{\een}{\end{equation*}}

\newcommand{\rd}{\mathcal}

\newcommand{\badat}{\begin{alignedat}}
\newcommand{\eadat}{\end{alignedat}}
\newcommand{\bitm}{\begin{itemize}}
\newcommand{\eitm}{\end{itemize}}
\newcommand{\bmat}{\begin{pmatrix}}
\newcommand{\emat}{\end{pmatrix}}
\newcommand{\ba}{\begin{align}}
\newcommand{\bas}{\begin{align*}}
\newcommand{\ea}{\end{align}}
\newcommand{\bse}{\begin{subequations}}
\newcommand{\ese}{\end{subequations}}
\newcommand{\gt}{\rightarrow}

\newcommand{\om}{\omega}
\newcommand{\ep}{\epsilon}
\newcommand{\virg}{\hspace{1 mm}, \hspace{8 mm}}
\newcommand{\pvirg}{\hspace{1 mm}; \hspace{8 mm}}

\begin{document}

%%\tableofcontents

%\begin{titlepage}
\begin{center}

\noindent{{\LARGE{Conserved charges in timelike Warped-AdS$_3$ spaces}}}

\smallskip
\smallskip

\smallskip
\smallskip
\smallskip
\smallskip
\noindent{\large{L. Donnay$^{1}$, J.J. Fern\'andez-Melgarejo$^{2}$, G. Giribet$^{1,3,4}$, A. Goya$^{3}$, E. Lavia$^{3,5}$}}

\smallskip
\smallskip

\end{center}

\smallskip
\smallskip
\centerline{$^1$ Universit\'{e} Libre de Bruxelles and International Solvay Institutes}
\centerline{{\it ULB-Campus Plaine CPO231, B-1050 Brussels, Belgium.}}

\smallskip
\smallskip
\centerline{$^2$ Center for the Fundamental Laws of Nature, Harvard University,}
\centerline{{\it Cambridge, MA 02138, USA.}}

\smallskip
\smallskip
\centerline{$^3$ Departamento de F\'{\i}sica, Universidad de Buenos Aires and IFIBA-CONICET}
\centerline{{\it Ciudad Universitaria, Pabell\'on 1, 1428, Buenos Aires, Argentina.}}

\smallskip
\smallskip
\centerline{$^4$ Instituto de F\'{\i}sica, Pontificia Universidad Cat\'{o}lica de
Valpara\'{\i}so}
\centerline{{\it Casilla 4059, Valpara\'{\i}so, Chile.}}

\smallskip
\smallskip
\centerline{$^5$ Argentinian Navy Research Office (DIIV), UNIDEF and CONICET}
\centerline{{\it Laprida 51, (1638) Vicente L\'opez, Buenos Aires, Argentina.}}

\bigskip

\bigskip

\bigskip

\bigskip

We consider the timelike version of Warped Anti-de Sitter
space (WAdS), which corresponds to the three-dimensional section of the G\"{o}del solution of four-dimensional cosmological Einstein equations. This
geometry presents closed timelike curves (CTCs), which are inherited from its
four-dimensional embedding. In three dimensions, this type of solutions can
be supported without matter provided the graviton acquires mass. Here, among the different ways to consistently give mass to the graviton in three dimensions, we consider the parity-even model known as New Massive Gravity (NMG). In
the bulk of timelike WAdS$_{3}$ space, we introduce defects that, from the three-dimensional
point of view, represent spinning massive particle-like 
objects. For this type of sources, we investigate the definition of quasi-local 
gravitational energy
as seen from infinity, far beyond the region where the CTCs appear. We also consider the covariant formalism applied to NMG to compute the mass and the angular momentum of spinning particle-like defects, and compare the result with the one obtained by means of the quasi-local stress-tensor. We apply these methods
to special limits in which the WAdS$_3$ solutions coincide with locally AdS$_3$ and locally AdS$_{2}\times \mathbb{R}$ spaces. Finally, we make some comments about the asymptotic symmetry algebra of asymptotically WAdS$_3$ spaces in NMG.  
\begin{equation*}
\end{equation*}

\section{Introduction}

In the last years, gravity about three-dimensional Warped Anti-de Sitter (WAdS$_{3}$) spaces has attracted attention
due to the fact that it represents one of the most interesting examples of
what has been dubbed \textit{non-AdS holography}. Different proposals
suggesting that quantum gravity in WAdS$_{3}$ space could be dual to a
two-dimensional theory with certain type of conformal invariance have
appeared in the literature \cite{warped, Song2, DHH, Hofman2}. It is therefore natural to
ask to what extent the holography-inspired techniques to compute observables such as conserved
charges can be extended to the case of timelike WAdS$_{3}$. The majority of the works
considering WAdS$_{3}$ holography in the literature are, however, concerned with the
spacelike WAdS$_{3}$ spaces. This is because, on the one hand, spacelike spaces can host black holes \cite{Clement}, which are particularly
interesting; on the other hand, the fact that stretched timelike
WAdS$_{3}$ spaces exhibit closed timelike curves (CTCs) is usually regarded as a pathology that makes this case
less physically sensible than its squashed spacelike analogue. Nevertheless, there are
still good reasons to study the definition of conserved charges in
asymptotically timelike WAdS$_{3}$ spaces. One such
motivation comes from dS/CFT: In dS/CFT, as originally proposed \cite{dSCFT}, the dual field theory is supposed
to be an Euclidean CFT formulated at future infinity. In the static patch,
the holographic picture is such that the dual CFT is located beyond the
cosmological horizon. Therefore, when trying to apply holographic
renormalization techniques, one has to propose a way to define the
regularized boundary stress-tensor far beyond the horizon, where the vector
that is timelike inside the static patch becomes spacelike. A particular
proposal to do so in dS space has been given in Ref. \cite{Minic}, where it was
proposed that conserved charges can be defined in terms of the holographic stress-tensor integrating on constant-$t$ codimension-2 surfaces, being $t$ the coordinate which is timelike inside the static patch. This proposal works well for dS/CFT and seems to be an
ingenious trick to deal with backgrounds that do not necessarily admit a
globally defined timelike Killing vector. A possible explanation of why the proposal in \cite{Minic} works has been recently given by Ref. {\cite{LD}}, where it has been shown that the dual CFT description does not necessarily have to be placed at the future conformal boundary, but also holds on any fixed timelike slice in the static patch. 

Here, we will raise the same kind of questions for timelike WAdS$_{3}$ spaces in New Massive Gravity (NMG) \cite{NMG}.  First, we will investigate to what extent we can provide
a notion of quasi-local energy in timelike WAdS$_{3}$ from far infinity. Secondly, we will study the conserved charges, such as mass and angular momentum of spinning particle-like objects in timelike WAdS$_3$ using the covariant formalism applied to NMG. We will see that, despite the fact that timelike WAdS$_3$ exhibits CTCs, this space admits a sensible definition of conserved charges. 

The main motivations to study the particular case of NMG are the following: Being a ghost free theory that propagates two degrees of freedom, NMG is the three-dimensional (3D) theory of gravity whose dynamics is closest to that of four-dimensional gravity, in the sense that it propagates two polarizations of a spin-2 particle (instead of zero as in 3D General Relativity, only one as in Topologically Massive Gravity (TMG), or three as in other 3D higher-curvature models). NMG is, on the other hand, the simplest model of three-dimensional gravity that preserves parity (unlike TMG, for instance). This latter property is of particular importance in the case of WAdS$_3$ backgrounds because it permits to distinguish between the {\it chiral features} arising from the $SL(2,\mathbb{R})\times U(1)$ symmetric geometry from chiral features inherent of parity-violating effects (which are present in theories such as TMG).

The paper is organized as follows: In Section 2, timelike WAdS$_3$ space is discussed and its main properties reviewed. In Section 3, we discuss the WAdS$_3$ spaces as solutions to three-dimensional massive gravity. In Section 4, we propose a definition of quasi-local gravitational energy for defects in asymptotically WAdS$_3$ spaces. We define the quasi-local stress-tensor for three-dimensional massive gravity and discuss the difficulties encountered when trying to compute both the mass and angular momentum of defects with this method. In addition, we compute the conserved charges associated to defects in timelike WAdS$_3$ in the covariant formalism adapted to massive gravity. We compare the results obtained for the timelike WAdS$_3$ defects with the computation of the mass and angular momentum of asymptotically spacelike WAdS$_3$ black holes. Section 5 contains our conclusions.

\section{Timelike WAdS$_{3}$ space}

Timelike WAdS$_3$ spaces are squashed or stretched deformations of asymptotically three-dimensional Anti-de Sitter spaces (AdS$_3$) \cite{Sandinista}. In the case of the stretched deformation, WAdS$_3$ corresponds
to the three-dimensional section of the Reboucas-Tiomno one-parameter
generalization \cite{reboucas, Spindel} of the G\"{o}del solution of
four-dimensional cosmological Einstein equations, and the existence of CTCs\ is a property inherited from its four-dimensional ancestor, the G%
\"{o}del universe \cite{Godel}. These spaces represent a workable example to address questions such as how to define physically sensible observables, such as conserved charges, in spaces with CTCs.

\subsection{Timelike WAdS$_{3}$ from G\"{o}del metric}

%\QTP{Body Math}
G\"{o}del cosmological solution is the direct product of the real line, $%
\mathbb{R}$, and a three-dimensional manifold $\Sigma $ equipped with a
metric \cite{Godel, haw-ellis} 
\begin{equation}
ds^{2}=-\left( d\hat{t}+e^{\sqrt{2}\omega x}dy\right) ^{2}+dx^{2}+\frac{1}{2}e^{2%
\sqrt{2}\omega x}dy^{2},  \label{godel_txy}
\end{equation}%
with coordinates $x,y,\hat{t} \in \mathbb{R}$, and $\omega $ being a real
parameter that represents the \textit{vorticity} of the G\"{o}del solution.
This coordinate system gives a complete chart of the space, and the
four-dimensional solution is then homeomorphic to $\mathbb{R}^{4}$. The
space is geodesically complete, and hence singularity free; it is spatially
homogeneous, though non-isotropic.

In a convenient system of coordinates, metric (\ref{godel_txy}) above takes the form%
\begin{equation}
ds^{2}=-\left( dt+\frac{2}{\omega }\sinh ^{2}\left( \frac{\omega
\rho }{\sqrt{2}}\right) d\phi \right) ^{2}+\frac{1}{2\omega ^{2}}\sinh ^{2}(\sqrt{2}%
\omega \rho )d\phi ^{2}+d\rho ^{2},  \label{E}
\end{equation}%
where the three-dimensional metric is now written as a Hopf fiber over the
hyperbolic plane. This space exhibits closed timelike curves, as it can be seen from the role played by coordinates $t$ and $\phi $ in the first term of (\ref{E}).

The prominent properties of the G\"{o}del space persist if one considers a
particular one-parameter deformation of the metric (\ref{E}) which, in
particular, permits to interpolate between the three-dimensional section of G%
\"{o}del space and AdS$_{3}$ \cite{reboucas}. This deformation is given by
the metric 
\begin{equation}
ds^{2}=-\left( dt+\frac{4\omega }{\lambda ^{2}}\sinh ^{2}\left( \frac{%
\lambda \rho }{2}\right) d\phi \right) ^{2}+\frac{\sinh ^{2}(\lambda \rho )}{%
\lambda ^{2}}d\phi ^{2}+d\rho ^{2},  \label{godelban}
\end{equation}%
which, apart from the vorticity $\omega $, includes an additional real
parameter $\lambda $ that controls the deformation. For the particular value 
$\lambda ^{2}=2\omega ^{2}$, metric (\ref{godelban}) corresponds to the
three-dimensional section of G\"{o}del solution (\ref{E}); when $\lambda ^{2}=4\omega ^{2}$ it corresponds to the universal
covering of AdS$_{3}$. For generic values of $\lambda $ and $\omega $ within
the range $0\leq \lambda ^{2}\leq 4\omega ^{2}$, metric (\ref{godelban})
describes the timelike stretched WAdS$_{3}$ spaces we will be concerned with.

It is convenient to consider a slightly different parameterization: Define
the parameter%
\begin{equation}
\ell ^{2}=\frac{2}{\lambda ^{2}-2\omega ^{2}},
\end{equation}%
and then use $\omega $ and $\ell ^{2}$ (instead of $\lambda $)\ to describe
the family of WAdS$_{3}$ metrics. For instance, in terms of $\omega $ and $%
\ell ^{2}$, the G\"{o}del solution corresponds to $\ell ^{2}=\infty $, while
AdS$_{3}$ space corresponds to $\ell ^{2}=\omega ^{-2}$. The range $0\leq
\lambda ^{2}\leq 4\omega ^{2}$, in terms of these parameters, translates
into $|\omega ^{2}\ell ^{2}|\geq 1$. Notice that $\omega ^{2}\ell ^{2} $ may take values between $-1$ and $-\infty $. Spaces with $|\omega ^{2}\ell ^{2}|< 1$ are also interesting, although present a different causal structure; they correspond to the timelike {\it squashed} WAdS$_3$ spaces.

Now, continuing with the convenient changes of coordinates, define the new
radial variable $r={2}{\lambda ^{-2}}\sinh^{2}( {\lambda \rho }/{2} )$, such that $r\in \mathbb{R}_{\geq 0}$. Metric (\ref{godelban}) now reads
\begin{equation}
ds^{2}=-dt^{2}-4\omega rdtd\phi +2\left( r+(\ell ^{-2}-\omega
^{2})r^{2}\right) d\phi ^{2}+\frac{dr^{2}}{2\left( r+(\ell ^{-2}+\omega
^{2})r^{2}\right) }.  \label{godel_trphi2}
\end{equation}

This is one of the standard ways of representing timelike WAdS$_{3}$ space.
The curvature invariants associated to this metric are constant, and take the
remarkably succinct form%
\begin{equation}
R_{\mu _{n}}^{\ \mu _{1}}R_{\mu _{1}}^{\ \mu _{2}}R_{\mu _{2}}^{\ \mu
_{3}}...R_{\mu _{n-1}}^{\ \mu _{n}}=(-1)^{n}\frac{2^{n}}{\ell ^{2n}}(\omega
^{2n}\ell ^{2n}+2).
\end{equation}

Another interesting property of metric (\ref{E}) is that it is spatially
homogeneous. As it happens with the universal covering of AdS, the WAdS spaces are
not globally hyperbolic.

The isometry group of WAdS$_3$ spaces (\ref{godel_trphi2}) is $SL(2,\mathbb{R})\times U(1)$, which is
generated by four out of the five Killing vectors that G\"{o}del solution
admits. This isometry is the 
remnant piece of the $SL(2,\mathbb{R})\times SL(2,\mathbb{R})$ isometry group of AdS$_3$ that survives through the stretched/squashed deformation.

From (\ref{godel_trphi2}), it is easy to verify that in the special point $\omega^2\ell^2=1$ the solution tends to AdS$_3$ space. Indeed, defining the new coordinates $\theta = t-\phi $ and $\rho^2=2r$ and replacing $\omega = \ell = 1$ in (\ref{godel_trphi2}), gives
\begin{equation}
ds^{2}_{\text{AdS}_3}=-(\rho^2+1)dt^{2}+\frac{d\rho^2}{(\rho^2+1)}+\rho^2d\theta^2.  \label{AdS3}
\end{equation}

\subsection{Introducing a defect}

Let us now introduce a pointlike defect in spacetime (\ref{godel_trphi2}).
This is achieved by performing the change 
\begin{equation}
\phi \rightarrow (1-\mu )\varphi ,\quad \text{with}\quad 0\leq \mu <1,
\label{acostex}
\end{equation}%
while keeping the same periodicity for the $\varphi $ coordinate, namely 
$\varphi \in \lbrack 0,2\pi )$. This certainly changes the global properties
of the space in a way that is equivalent to introducing an angular deficit $%
\delta \phi =\mu /(2\pi )$ in the original angular coordinate.\ By doing (%
\ref{acostex})\ and rescaling the radial coordinate as $r\rightarrow
r/(1-\mu )$ one finds the metric 
\begin{eqnarray}
ds^{2} &=&-dt^{2}-4\omega rdtd\varphi +2r\left( (\ell ^{-2}-\omega
^{2})r+(1-\mu )\right) d\varphi ^{2}  \notag \\
&&+\frac{dr^{2}}{2r\left( (\omega ^{2}+\ell ^{-2})r+(1-\mu )\right) },
\label{godel_cosmones2}
\end{eqnarray}%
where $t\in \mathbb{R}$, $r\in \mathbb{R}_{\geq 0}$, and $\varphi \in 
\mathbb{[}0,2\pi \mathbb{)}$. This metric shares the asymptotic behavior
with (\ref{godel_trphi2});\ namely both have the large $r$ behavior%
\begin{equation}
ds^{2}=-dt^{2}-4\omega rdtd\varphi +2(\ell ^{-2}-\omega ^{2})r^{2}d\varphi ^{2}+%
\frac{dr^{2}}{2r^{2}(\ell ^{-2}+\omega ^{2})}+h_{\mu \nu }dx^{\mu }dx^{\nu },
\label{once}
\end{equation}%
with, in particular, $\delta g_{\varphi \varphi }\equiv h_{\varphi \varphi }\simeq \mathcal{O}(r)\ $and $\delta g_{rr}\equiv
h_{rr}\simeq \mathcal{O}(r^{-3})$.

Metric (\ref{godel_cosmones2}) represents a particle-like object located at $r=0$, in the bulk of G\"odel universe. The object disappears when $\mu$ tends to 
zero, which permits to anticipate that $\mu$ is somehow related to the mass of the defect. More general defects will be introduced later (see (\ref{godel}) 
below), which will represent spinning point particles in G\"odel spacetime. 

\section{Timelike WAdS$_{3}$ space in massive gravity}

\subsection{WAdS$_3$ spaces as gravity backgrounds}

A feature that makes WAdS$_{3}$ spaces of particular interest is that these
geometries appear as exact solutions of a large variety of models, including
string theory \cite{inta, instring}, topologically massive gauge theories \cite{intopologicallygaugetheories, israel, yotromas, 3dorigin}, higher-derivative
theories \cite{Clem}, bi-gravity theories \cite{inbigravity}, and Einstein gravity non-minimally coupled to
matter fields \cite{inEinstein}. A minimal setup in which WAdS$_{3}$ spaces appear is three-dimensional gravity with no matter fields. Indeed, spacelike and
timelike WAdS$_{3}$ geometries are exact solutions of pure three-dimensional
gravity provided one gives a small mass to the graviton. The graviton mass
is what ultimately induces the \textit{vorticity} required to support the G\"{o}del universe or, more precisely, the three-dimensional non-trivial part
of it. In three-dimensions, there are different manners to give mass to the
graviton in a consistent way. Here, we will adopt the particular parity-even
theory of massive gravity proposed in Ref. \cite{NMG}, usually called New Massive Gravity (NMG), which we will review in the next subsection. Our method to compute the
quasi-local gravitational energy, in Section 4, amounts to define a boundary stress-tensor for NMG,
which is the generalization of the Brown-York quasi-local stress-tensor. For NMG theory, such a tensor exists and has been defined in Ref. \cite{Hohm-Tonni}. We will consider such a definition of the quasi-local stress-tensor
and use it to compute the mass of the defect in timelike WAdS$_{3}$ as seen
from infinity, i.e. from the region that is beyond the radius where CTCs
appear. 

We will first consider a defect in timelike WAdS$_{3}$ space, which comes
to represent a massive spinless pointlike object. From the four-dimensional point of
view, this is like considering a local cosmic string in the G\"{o}del universe.
We will propose a physically sensible definition of mass for such a
highly-localized source. Intriguingly, the result we will obtain will be shown to account for $1/2$ of the Arnowitt-Deser-Misner (ADM) mass of the defect. In addition, the definition of charges in terms of the quasi-local stress-tensor will prove to be not suitable to compute the angular momentum of spinning defects, the failure being associated to the impossibility of regularizing the boundary stress-tensor by means of local counterterms. This will eventually lead us to consider an alternative approach to compute charges. We will consider, in Section 5, the covariant formalism for computing charges in NMG. Let us now introduce the theory.

\subsection{Three-dimensional New Massive Gravity}

In this section, we will discuss asymptotically timelike WAdS$_{3}$ spaces in
the specific context of three-dimensional NMG. The action of
the theory consists of three distinct contributions, namely%
\begin{equation}
S=S_{\text{EH}}+S_{\text{NMG}}+S_{\text{B}},  \label{action}
\end{equation}%
where the first term is the Einstein-Hilbert action 
\begin{equation}
S_{\text{EH}}=\frac{1}{16\pi G}\int_{\Sigma }d^{3}x\sqrt{-g}\left( \sigma
R-2\Lambda \right) ,  \label{pupo}
\end{equation}%
with $\sigma =\pm 1$ being a sign that effectively controls the sign of the
Newton constant. The second term in (\ref{action})\ is given by
\begin{equation}
S_{\text{NMG}}=\dfrac{1}{16\pi Gm^{2}}\int_{\Sigma }{d^{3}x\sqrt{-g}}\left( {%
R_{\mu \nu }R^{\mu \nu }-\dfrac{3}{8}R^{2}}\right) ,  \label{pupa}
\end{equation}
where $m$ is the mass of the graviton. The third term in (\ref{action}) is the boundary action, needed for the
variational principle to be well-posed. We will discuss the boundary action
in the next subsection.

Let us recall the main properties of theory (\ref{pupo})-(\ref{pupa}): Around maximally symmetric backgrounds, its linearized limit coincides with the massive spin-$2$ 
Fierz-Pauli action, representing a fully covariant extension of the latter. At a generic point of the parameter space ($\Lambda $, $m$), the theory propagates two 
massive local degrees of freedom. In addition, NMG admits a rich set of solutions, such as Scr\"odinger invariant spaces \cite{Sch}, Lifshitz spaces and Lifshitz black holes \cite{Lif}, logarithmic deformation of the Ba\~nados-Teitelboim-Zanelli geometry \cite{Clemooo}, hairy (A)dS$_3$ black holes \cite{Troncoso}, WAdS$_3$ black holes, and others \cite{Clem, Siampos}.

The equations of motion derived from (\ref{pupo}) and (\ref{pupa}) are
\begin{equation}
R_{\mu \nu }-\frac{1}{2}Rg_{\mu \nu }+\sigma \Lambda g_{\mu \nu }+\frac{%
\sigma }{2m^{2}}K_{\mu \nu }=0 ,  \label{eom}
\end{equation}%
which, apart from the Einstein tensor, involve the tensor%
\begin{equation*}
K_{\mu \nu }=2\square {R}_{\mu \nu }-\frac{1}{2}\nabla _{\mu }\nabla _{\nu }{%
R}-\frac{1}{2}\square {R}g_{\mu \nu }+4R_{\mu \alpha \nu \beta }R^{\alpha
\beta }-\frac{3}{2}RR_{\mu \nu }-R_{\alpha \beta }R^{\alpha \beta }g_{\mu
\nu }+\frac{3}{8}R^{2}g_{\mu \nu }.
\end{equation*}

Timelike WAdS$_{3}$ metrics (\ref{godel_cosmones2}) solve the equations of
motion (\ref{eom}) provided the coupling constants satisfy\footnote{These solutions persist if one introduces in the equations of motion the Cotton tensor of TMG.}
\begin{equation}
\Lambda =-\dfrac{(11\omega ^{4}\ell ^{4}+28\omega ^{2}\ell ^{2}-4)\sigma }{%
2(19\omega ^{2}\ell ^{2}-2)\ell ^{2}}\,,\quad \quad m^{2}=-\dfrac{(19\omega
^{2}\ell ^{2}-2)\sigma }{2\ell ^{2}}\,. \label{20}
\end{equation}%
Recall that AdS$_{3}$ space corresponds to $\omega ^{2}\ell ^{2}=1$, for
which $\Lambda =-35\sigma /(34\ell ^{2})$ and $m^{2}=-17\sigma /(2\ell ^{2})$.

\subsection{Boundary terms}

To discuss boundary terms $S_{\text{B}}$, let us first rewrite (\ref{pupa}) as follows:
\begin{equation}
S_{\text{NMG}}=\frac{1}{16\pi G}\int_{\Sigma }{d^{3}x\sqrt{-g}}\left( {%
f^{\mu \nu }(R_{\mu \nu }-\frac{1}{2}Rg_{\mu \nu })-\frac{1}{4}m^{2}(f_{\mu
\nu }f^{\mu \nu }-f^{2})}\right) .  \label{SNMG}
\end{equation}%
This includes an auxiliary field $f_{\mu \nu }$, represented by a rank-2
symmetric tensor. After varying with respect to $f_{\mu \nu }$, one finds%
\begin{equation}
f_{\mu \nu }=\dfrac{2}{m^{2}}(R_{\mu \nu }-\dfrac{1}{4}Rg_{\mu \nu }),
\label{f}
\end{equation}%
which can be plugged back into (\ref{SNMG}) to reproduce the higher-curvature
term (\ref{pupa}).

The next step is to consider the ADM type decomposition in the
radial direction; that is 
\begin{equation}
ds^{2}=N^{2}dr^{2}+\gamma _{ij}(dx^{i}+N^{i}dr)(dx^{j}+N^{j}dr),  \label{ADM}
\end{equation}%
where $N^{2}$ is the radial analogue of the lapse function, $N^{i}$ are the shift functions and $\gamma
_{ij} $ is the two-dimensional metric induced on the constant-$r$ surfaces.
The Latin indices $i,j=0,1$, refer to the coordinates on the constant-$r$
surfaces (namely $x^{0}=t$, $x^{1}=\varphi $), while the Greek indices $\mu
,\nu =0,1,2$, refer to all coordinates, including the radial direction $%
x^{2}=r$.

Boundary terms $S_{\text{B}}$ are introduced in (\ref{action}) for the
variational principle to be defined in such a way that both the metric $%
g_{\mu \nu }$ and the auxiliary field $f_{\mu \nu }$ are fixed on the
boundary $\partial \Sigma $; see \cite{Hohm-Tonni} for details. The boundary
action is then given by 
\begin{equation}
S_{\text{B}}=-\frac{1}{8\pi G}\int_{\partial \Sigma }d^{2}x\sqrt{-\gamma }%
\left( K+\frac{1}{2}\hat{f}^{ij}(K_{ij}-\gamma _{ij}K)\right) ,  \label{Sb}
\end{equation}%
where, as said, $\gamma _{ij}$ is the metric induced on $\partial \Sigma $,  $\gamma =\det (\gamma _{ij})$
and $K_{ij}$ is the extrinsic curvature, with $K=\gamma ^{ij}K_{ij}$. $\hat{f%
}^{ij}$ in (\ref{Sb})\ comes from decomposing the auxiliary field $f^{\mu
\nu }$ as follows $ f^{\mu \nu }=\delta _{i}^{\mu }\delta _{j}^{\nu }f^{ij}+2\delta _{r}^{(\mu
}\delta _{i}^{\nu )}h^{i}+\delta _{r}^{\mu }\delta _{r}^{\nu }s $ and then defining $ \hat{f}^{ij}\equiv f^{ij}+2h^{(i}N^{j)}+sN^{i}N^{j},$ and $\hat{f}\equiv
\gamma _{ij}\hat{f}^{ij}$, where $a^{(\mu }b^{\nu )}\equiv (a^{\mu }b^{\nu }+a^{\nu }b^{\mu })/2$.

The first term in (\ref{Sb}) corresponds to the Gibbons-Hawking term of
General Relativity, while the other two terms come from the higher-curvature
terms of (\ref{SNMG}). These terms are preliminary elements to define the boundary stress-tensor, which we will discuss in the next section.

\section{Conserved charges}

\subsection{The quasi-local stress-tensor}

The Brown-York quasi-local stress-tensor $T_{ij}$ is obtained by varying action (\ref%
{action}) with respect to the metric $\gamma ^{ij}$, \cite{BY}. That is,%
\begin{equation}
T_{ij}=\frac{2}{\sqrt{-\gamma }}\frac{\delta S}{\delta \gamma ^{ij}}_{|r=%
\text{const}},  \label{Tij}
\end{equation}%
which yields \cite{Hohm-Tonni} 
\begin{eqnarray}
T^{ij} &=&\frac{1}{8\pi G}(K^{ij}-K\gamma ^{ij})-\frac{1}{8\pi G}\left( 
\frac{1}{2}\hat{f}K^{ij}+\nabla ^{(i}\hat{h}^{j)}-\frac{1}{2}\nabla _{r}\hat{%
f}^{ij}+K_{k}^{(i}\hat{f}^{j)k}\right.  \notag \\
&&\left. -\frac{1}{2}N^{2}sK^{ij}-\gamma ^{ij}(\nabla _{k}\hat{h}^{k}-\frac{1%
}{2}N^{2}sK+\frac{1}{2}\hat{f}K-\frac{1}{2}\nabla _{r}\hat{f})\right),
\label{OTij}
\end{eqnarray}%
where $\hat{h^{i}}=N(h^{i}+sN^{i}N^{j})$. The covariant $r$-derivative $%
\nabla _{r}$ acting on $\hat{f}^{ij}$ is defined as follows 
\begin{equation}
\nabla _{r}\hat{f}^{ij}=\frac{1}{N}\left( \partial _{r}\hat{f}%
^{ij}-N^{k}\partial _{k}\hat{f}^{ij}+2\hat{f}^{k(i}\partial
_{k}N^{j)}\right) ,\quad \quad \nabla _{r}\hat{f}=\frac{1}{N}\left( \partial
_{r}\hat{f}-N^{k}\partial _{k}\hat{f}\right) .  \label{OOTij}
\end{equation}

When taking the limit $r\rightarrow \infty $ in the definition (\ref{Tij}),
stress-tensor (\ref{OTij}) is found to diverge. Without a proper regularization procedure, this would result in an
infinite value for the conserved charges. To solve this problem, one may try
to improve the definition (\ref{Tij}) by including additional boundary terms
to the action, provided such terms do not spoil the
variational principle. In Ref. \cite{GiribetGoya}, this method was applied
to the case of spacelike WAdS$_{3}$. It was shown that, despite the persistent divergences
of some components of $T_{ij}$, adding a boundary
cosmological constant term to $S_{\text{B}}$ makes the functional
action finite and yields a finite quasi-local energy. We can try to do the same here
for the timelike case and improve the stress-tensor (\ref{Tij}) by adding a piece%
\begin{equation}
T_{ij}\to T_{ij}-\frac{\zeta }{8\pi G} \gamma _{ij}  \label{nn},
\end{equation}%
which would come from a boundary contribution%
\begin{equation}
S_{\text{B}} \to S_{\text{B}}+\frac{\zeta }{8\pi G} \int d^{2}x\sqrt{-\gamma },
\label{mm}
\end{equation}%
where $\zeta $ is a coefficient fixed by requiring the action to be finite. The value of this coefficient is found to be
\begin{equation}
\zeta  = -\frac{\sigma 8\omega^2\ell \sqrt{2(\omega^2\ell^2+1)}}{(19\omega^2\ell^2-2)}.
\end{equation}

\subsection{Quasi-local gravitational energy}

The boundary stress-tensor (\ref{Tij}), once improved by the adding of (\ref%
{nn}), yields the definition of conserved charges $Q_{\tilde{\xi }}$,
associated to vectors $\tilde{\xi}$ that generate isometries on $\partial
\Sigma $. These {\it boundary Killing vectors} $\tilde{\xi}$ are defined by the equation
\begin{equation}
\pounds_{\tilde{\xi}}\gamma _{ij}=0\,,  \label{Uma}
\end{equation}%
for the induced metric. Then, the charges are defined by integrating
the projection of the boundary stress-tensor on the vector $\tilde{\xi}$ and a
unitary vector $u$ that is orthogonal to the constant-$t$
surfaces. That is,%
\begin{equation}
Q_{\tilde{\xi} } =\int  d\varphi \ \varrho \ u^{i} T_{ij}\tilde{\xi}%
^{j},  \label{result}
\end{equation}%
where $\varrho $ is given by the induced metric written in the form%
\begin{equation}
d\Sigma ^{2}=-N_{\Sigma }^{2}dt^{2}+\varrho ^{2}(dt+N_{\Sigma }^{\varphi }d\varphi
)^{2} .  \label{ADM-t}
\end{equation}%

In particular, for the WAdS$_3$ defects we have 
\be
\badat{3}
\varrho ^{2}& =2(1-\mu )r+2(\ell ^{-2}-\omega ^{2})r^{2}\,, \\
N_{\Sigma }^{\varphi }& =-\dfrac{\omega r}{(1-\mu )r+(\ell ^{-2}-\omega ^{2})r^{2}}%
\,, \\
N_{\Sigma }^{2}& =1+\dfrac{2\omega ^{2}r^{2}}{(1-\mu )r+(\ell ^{-2}-\omega
^{2})r^{2}}\,.
\eadat
\ee%

With these ingredients, we are ready to compute the mass of the defects: The unitary vector orthogonal to the constant-$t$ surfaces is given by $u=-N_{\Sigma
}(r)dt$. Considering a timelike boundary Killing vector 
\begin{equation}
	\tilde{\xi}^i = N_{\Sigma} u^i \,,
	\label{Bdy.KV}
\end{equation}
(i.e. timelike in the region where the source is located) we find a value for the quasi-local energy $\mathcal{M}= Q_{\tilde{\xi} }$, which reads  
\begin{equation}
\mathcal{M}  =\dfrac{2\,\sigma \,\omega ^{2}\ell ^{2}\,(\mu -1)}{(19\omega
^{2}\ell ^{2}-2)G}=\frac{2(\mu -1)}{19G}\left( \sigma -\dfrac{1}{m^{2}\ell
^{2}}\right) ,  \label{AA}
\end{equation}
where we used that $2m^{2}\ell ^{2}\sigma =2-19\omega ^{2}\ell ^{2}$. 

Let us first compare the result (\ref{AA}) with the special case of locally AdS$_{3}$ solutions, which correspond to $\omega
^{2}\ell ^{2}=1$. In this case, (\ref{AA}) reduces to 
\begin{equation}
\mathcal{M}_{\omega^2\ell^2=1}=\frac{2\sigma (\mu -1)}{17G},  \label{JI}
\end{equation}%
and, indeed, this is seen to match the mass of a defect in locally AdS$%
_{3}$ space in NMG. To see this explicitly, let us be reminded of the fact that
in the case of NMG\ in AdS$_{3}$ the mass of a deficit angle (a particular
case of the BTZ geometry)\ is given by \cite{NMG2}%
\begin{equation}
\mathcal{M}_{\text{AdS}_3}=\frac{(\mu -1)}{8G}\left( \sigma +\frac{1}{%
2m^{2}\ell ^{2}}\right) =\frac{2\sigma (\mu -1)}{17G},  \label{29bis}
\end{equation}%
where we used that $\omega ^{2}\ell ^{2}=1$ precisely corresponds to $%
2m^{2}\ell ^{2}\sigma =-17$. That is, (\ref{AA}) reduces to the
value \eqref{JI} at that point of the parameter space. In principle, we could be tempted to take this matching as a consistency check of the result (\ref{AA}). However, if we think of it carefully, we conclude that there is a priori no good reason to expect (\ref{AA}) to coincide with (\ref{29bis}) in the $\omega^2\ell^2 \to 1$ limit. This is because, even when in that limit WAdS$_3$ space becomes AdS$_3$ space, the latter shows up in a coordinate system which is not the one usually considered when computing the ADM charges of BTZ geometry. This is similar to what happens in the case of asymptotically WAdS$_3$ black holes, whose conserved charges, as functions of the horizons radii, do not tend to the charges of BTZ black holes in the $\nu \to 1$ limit (being $\nu$ the parameter that controls the deformation in that case; see the conventions in \cite{warped}). In fact, we will see in the next section that the correct value of the gravitational mass associated to a pointlike defect in timelike WAdS$_3$ space coincides with (\ref{AA}) only up to a factor of $1/2$. This feature has already been observed in the context of spacelike WAdS$_3$ solutions \cite{GiribetGoya}.

As it happens with spacelike WAdS$_3$ spaces, the method of computing charges using the quasi-local stress-tensor (\ref{Tij}) does not suffice to give a finite result for the angular momentum of spinning defects. This is basically because there seems to be no manner to regularize all the components of (\ref{Tij}) by means of local boundary counterterms. This means that, in order to study spinning defects, it is necessary to consider a different method for computing conserved charges. With this motivation, we will consider in the following section the covariant formalism. 

\subsection{Covariant formalism in New Massive Gravity}

Let us now consider spinning defects. The metric of G\"{o}del spacetime with both mass and angular momentum reads \footnote{Notice that we can assign dimensions to the parameters and coordinates as follows: $[t]=l^1, [r]=l^2, [\phi]=l^0, [\ell]=l^1, [\om]=l^{-1}, [\mu]= l^0, [j]=l^0$, where $l$ has dimension of length.}
\begin{equation}
	d{s}^2 =-dt^2  -4\om r dt d\varphi +  \frac{dr^2}{\left( 2r^2\omega^2 +\lambda_{\mu , j} ( r) \right)}- \left( 2r^2\omega^2 -\lambda_{\mu , j} ( r) \right)d\varphi^2 ,
\label{godel}
\end{equation}
where 
\begin{equation}
\lambda_{\mu , j} ( r) = \frac{2r^2}{\ell^2} +2 (1-\mu)r -j \ell^2, 
\end{equation}
and where $t \in \mathbb R$, $r \in \mathbb R_{\geq 0}$, $0\leq \mu \leq 1$, and $\phi \in [0, 2\pi)$. Metric (\ref{godel}) involves a new parameter $j \in \mathbb R$, and reduces to (\ref{godel_cosmones2}) when $j=0$. Notice also that only $\xi^{t}\sim \partial_t$ and $\xi^{\varphi }\sim \partial_{\varphi }$ out of the four generators of $SL(2, \mathbb{R}) \times U(1)$ survive as exact Killing vectors of the metric \eqref{godel}.

Such as in the case of the parameter $\mu $, the introduction of $j$ is achieved by means of a (improper, i.e. not globally well-defined) diffeomorphism from metric (\ref{godel_trphi2}). Metric \eqref{godel} solves the equations of motion (\ref{eom}) for the parameters (\ref{20}).
    
In the covariant formalism \cite{Barnich:2001jy,Barnich:2007bf}, conserved charges associated to an asymptotic Killing vector $\xi$ are given in three spacetime dimensions by the expression
%\footnote{
%The form $k_\xi$ in \eqref{bms3:kxi} can be defined in terms of a symplectic form by $dk_\xi=W$ with the symplectic form $W$ defined by $W=\frac{1}{2}I_{\delta\phi}^n\left(\frac{\delta L}{\delta\phi}\delta\phi\right)$, where $I_{\delta\phi}^n$ is the homotopy operator acting on an $n$ form. This expression of $k_\xi$ differs from the one proposed by Iyer-Wald \cite{Iyer:1994ys}, $dk_\xi^{IW}=\omega^{IW}$, with the symplectic form $\omega^{IW}$ defined by $\omega^{IW}=\delta (I_{\delta\phi}^nL)$. The two different symplectic forms differ by a total derivative term, $W=\omega^{IW}+dE$ with $E=\frac{1}{2}(I_{\delta\phi}^{n-1}I_{\delta\phi}^nL)$. This $E$ term plays no role in the case of exact symmetries but is however relevant in the asymptotic context. See \cite{Barnich:2007bf} appendix A5 for more details.} 
\be
\delta Q_\xi [\delta g, g]=\frac{1}{16 \pi G} \int_{0}^{2\pi} \sqrt{-g}\, \ep_{\mu \nu \varphi} \,k_\xi^{\mu \nu}[\delta g, g]d\varphi,
\label{formulacharge}
\ee
with $g$ a solution, $\delta g$ a linearized perturbation around it, and $k_\xi^{\mu \nu}[\delta g, g]$ being a one-form potential of the linearized theory. In \cite{Nam}, this potential was computed for exact Killing vectors in NMG using the so-called Abbott-Deser-Tekin (ADT) formalism. The result can be written
\be
k_{\xi }^{\mu \nu}=Q^{\mu \nu}_R+\frac{1}{2m^2}Q^{\mu \nu}_{K},
\label{kexact}
\ee
where the first contribution comes from the pure GR part of the equations of motion, while $Q^{\mu \nu}_{K}$ accounts for the contribution of the $K_{\mu \nu}$ tensor of NMG, whose explicit expression can be found in equations (22), (28) and (29) in \cite{Nam}, respectively.

\subsection{Mass and angular momentum in the covariant formalism}

One can use \eqref{kexact} and plug it into \eqref{formulacharge} to compute the (variation of the) mass and angular momentum, for which the Killing vectors are, respectively, $\partial_t$ and $\partial_\varphi$. This procedure has been implemented in a Mathematica code and, for $\sigma =1$, gives\footnote{This result is up to $\mu $-independent and $j$-independent terms, which can not be gathered in the integration.}
\be
\rd M =\frac{4 (\mu -1) \ell^2  \om^2}{G(19 \ell^2 \om^2-2)}, 
\label{MGod}
\ee
and
\be
\rd J =-\frac{4 j \ell^4 \om^3}{G(19 \ell^2 \om^2-2)},
\label{JGod}
\ee
for the mass and the angular momentum of the solution \eqref{godel}, respectively. Notice that, as expected, the angular momentum changes its sign when $\omega $ does so.\\

Expressions \eqref{MGod}, \eqref{JGod} are the correct values of the conserved charges. Intriguingly, the Brown-York quasi-local energy obtained in (\ref{AA}) gives only one half of the mass \footnote{The same phenomenon has been observed in the case of spacelike black holes, where the quasi-local energy gives one half of the black hole mass computed by covariant methods, the latter being the value fulfilling the first principle \cite{Nam}.}.

A special case to consider is the actual G\"odel spacetime, which corresponds to the limit $\ell \to \infty $. In this case, the mass formula (\ref{MGod}) yields
\begin{equation}
\mathcal{M}_{\text{G\"od}}=\frac{4 (\mu -1)}{19G},  \label{29bisa}
\end{equation}
which is independent of $\omega $. For $\mu =0$ the result is negative and is of crucial importance in the study of the spacelike WAdS$_3$ black hole spectrum \cite{DHH,LG}.

Another special case to analyze is the AdS$_{2}\times \mathbb{R}$ space. This corresponds to the limit $\omega \to 0$. To see this explicitly, we define coordinate $\tilde{\rho}^2 = 1+4(r^2/\ell^4+r/\ell^2)$, in which the metric for $ \omega=0$ takes the form
\begin{equation}
ds^{2}_{|\omega=0}=-dt^2+ds^{2}_{\text{AdS}_2}=-dt^2+\frac{\ell^2}{2}(\tilde{\rho}^2-1)d\phi^2+\frac{\ell^2}{2}\frac{d\tilde{\rho}^2}{(\tilde{\rho}^2-1)}.  \label{AdS2}
\end{equation}

In this case, the mass also tends to zero, 
\begin{equation}
\mathcal{M}_{\mathbb{R}\times \text{AdS}_2}=0. 
\end{equation}

Locally AdS$_{2}\times \mathbb{R}$ spaces appear in the limit in which (\ref{20}) yields $\Lambda=-m^2$ \cite{Troncoso}.

\section{Conclusions}

In this paper, we have investigated the definition of conserved charges in timelike WAdS$_3$ spacetimes, which exhibit CTCs. We have considered these spaces in the context of NMG. Timelike WAdS$_3$ spacetimes in NMG represent a workable example to address questions such as how to define physically sensible observables, such as conserved charges, in spaces that do not possess a globally defined timelike Killing vector.

For stretched and squashed timelike WAdS$_3$ spaces, we have investigated several features related to the feasibility of defining conserved charges. One of the questions we have addressed was how to provide a sensible definition of quasi-local gravitational energy
in these spacetimes that exhibit CTCs. The motivation for doing this was studying to what extent the holography-inspired methods can be applied to this example of non-AdS holography. {We have succeeded in doing this for non-spinning defects. However, the difficulties encountered when trying to adapt this method to spinning solutions eventually led us to consider an alternative way of computing charges. We have resorted to the covariant formalism applied to NMG, which was shown to be suitable to compute the mass and angular momentum of a more general type of defects that represent spinning particle-like objects in the bulk of WAdS$_3$.}

The question remains as to whether it is possible to formulate a holographic renormalization recipe in WAdS$_3$ spaces. The obstruction encountered when trying to do this in Section 4 was the impossibility of regularizing the full boundary stress-tensor in terms of local boundary counterterms. This phenomenon had also been observed both in TMG and in NMG for the case of spacelike WAdS$_3$, suggesting this is a general feature of this type of backgrounds. Whether or not this problem is related to the lack of Lorentz invariance in the dual theory is still to be understood. 

Before concluding, let us mention that the covariant method for computing charges discussed in this paper can be adapted to the case of charges associated to asymptotic isometries. In a companion paper \cite{LG}, it will be shown that the algebra of charges in asymptotically WAdS$_3$ spaces is given by an infinite-dimensional algebra that coincides with the semidirect sum of Virasoro algebra with non-vanishing central charge and an affine $\hat{u}(1)_k$ Ka\v{c}-Moody algebra.

\section*{Acknowledgments}

The authors thank Geoffrey Comp\`{e}re, St\'ephane Detournay, C\'edric Troessaert for discussions. This work has been partially funded by FNRS-Belgium (convention FRFC PDR T.1025.14 and convention IISN 4.4503.15), by the Communaut\'{e} Fran\c{c}aise de Belgique
through the ARC program and by a donation from the Solvay family. L.D.~is a research fellow of the ``Fonds pour la Formation \`a
la Recherche dans l'Industrie et dans l'Agriculture''-FRIA
Belgium and her work is supported in part by IISN-Belgium and by
``Communaut\'e fran\c caise de Belgique - Actions de Recherche
Concert\'ee. The work of J.F-M. has been supported by a Postdoctoral Scholarship and Fundaci\'{o}n S\'{e}neca - Talento Investigador Program. The support of CONICET, FNRS+MINCyT and UBA through grants PIP 0595/13, BE 13/03 and UBACyT 20020120100154BA, respectively, is greatly acknowledged. 

\section*{Appendix: Relation with the WAdS$_3$ black holes}

As we will see, the relation between timelike charges we have obtained and the mass and angular momentum of the so-called Warped black holes (WBH) is not as simple as one could a priori think. WBHs are black hole solutions that asymptote stretched spacelike WAdS$_3$ space; see \cite{warped} and references therein. As we will describe below, these black holes can be obtained from the timelike solution by means of a complex change of coordinates: Consider first the double Wick rotation
\begin{eqnarray}
t \gt i\tau \virg \varphi \gt -i\Theta \virg \om \gt-\om \virg r\gt -r \virg j\gt -j ,  \label{i} 
\end{eqnarray}
and, secondly, $\tau =t'-\ell\sqrt j \Theta .$ Finally, in order to compare with the coordinates used in the literature, let us rescale time as $t'\gt L T .$

The change of coordinates above maps the timelike metric \eqref{godel} into the WBH solution\footnote{See Eq. (4.1) in Ref. \cite{warped}.}
\be
\badat{2}
ds^2= L^2 dT^2 + \frac{L^2 \, dR^2}{(\nu^2+3)(R-r_+)(R-r_-)}+L^2 (2\nu R -\sqrt{r_+r_-(\nu^2+3)})dTd\Theta\\
+\frac{R L^2}{4}\left[3(\nu^2-1)R+(\nu^2+3)(r_++r_-)-4\nu \sqrt{r_+r_-(\nu^2+3)}\right]d\Theta^2,
\label{warped}
\eadat
\ee
with $R=-2r/L^2$ and provided one identifies the parameters as follows
\be
\badat{2}
&\nu =\om L \pvirg L^2=\frac{3}{\om^2 +2 \ell^{-2}} ;\\
& r_{\pm}=\frac{\ell^2}{L^2}\left[\frac{-(1-\mu)\pm \sqrt{(1-\mu)^2-2(\om^2 \ell^2+1)j}}{(\om^2 \ell^2+1)}\right].
\label{mapping}
\eadat
\ee
Notice the useful relations
\be
r_+ + r_-= \frac{2\ell^2(\mu-1)}{L^2(1+\ell^2 \om^2)} \pvirg r_+ r_-= \frac{2j\ell^4}{L^4(1+\ell^2 \om^2)}.
\ee
The timelike and spacelike Killing vectors are related in the following way
\be
\partial_t=\frac{i}{L}\partial_T \virg \partial_\varphi=\frac{\ell}{L}\sqrt j \partial_T+i\partial_\Theta.
\ee

This charge dependent change of coordinates makes the relation between timelike and spacelike charges more involved than a mere analytic continuation.

Changing in \eqref{warped} $LT \gt t$, $R \gt r$ and $L\Theta \gt \varphi$, we can assign the dimensions as $[t]=l^1, [r]=l^1, [\varphi]=l^0, [L]=l^1, [\nu]=l^0, [r_\pm]=l^1$ and the expression of the mass of the WBH then becomes\footnote{This expression comes from (D.4) in \cite{Nam}, which coincides with (27) in \cite{GiribetGoya} without the extra factor $1/2$ which should be absent. Note that this expression has also been obtained independently with the covariant formalism.}
\be 
\rd M_\text{WBH}=Q_{\partial_T}=\frac{\nu (\nu^2+3)}{G L(20 \nu^2 -3)} \left((r_-+r_+)\nu - \sqrt{r_+r_-(\nu^2+3)}\right), 
\ee
while the expression for the angular momentum is \footnote{Result taken from (30) in \cite{GiribetGoya} which has been crossed checked with \cite{Clem}. Note that this expression has also been obtained independently in the covariant formalism.} 
\be
\rd J_\text{WBH}=Q_{\partial_{\Theta }}=\frac{\nu (\nu^2+3)}{4G L(20 \nu^2 -3)} \left((5\nu^2 +3)r_+r_- -2\nu \sqrt{r_+r_-(\nu^2+3)}(r_++r_-)\right).
\ee

Using the relations \eqref{mapping} between the spacelike and timelike parameters, one observes that going from the timelike to the spacelike metric involves a charge-dependent and globally not-well defined change of coordinates, namely the definition $\tau =t'-\ell\sqrt j \Theta $ above. This implies that the spacelike and timelike charges do not coincide. Only in the case $j=0$, one sees that the masses are related\footnote{Up to a $\mu$-independent factor, which can not seen in the integration.} according to $\partial_t \sim L^{-1}\partial_T$,
\be
\rd M_\text{WBH}|_{j=0}={L^{-1}}\rd M .
\ee

It is important to remark that, in the case of spinning defects in timelike WAdS$_3$, and due to the $j$-dependent change of coordinates, the conserved charges can not be simply obtained from the mass and angular momentum of spacelike solutions. 

\providecommand{\href}[2]{#2}\begingroup\raggedright\endgroup


\begin{thebibliography}{10}



\bibitem{warped}
D.~Anninos, W.~Li, M.~Padi, W.~Song, and A.~Strominger, ``{Warped AdS(3) Black
  Holes},'' \href{http://dx.doi.org/10.1088/1126-6708/2009/03/130}{{\em JHEP}
  {\bf 0903} (2009)  130},
\href{http://arxiv.org/abs/0807.3040}{{\tt arXiv:0807.3040 [hep-th]}}.
%%CITATION = ARXIV:0807.3040;%%.

\bibitem{Song2}
W.~Song and A.~Strominger, ``{Warped AdS3/Dipole-CFT Duality},''
  \href{http://dx.doi.org/10.1007/JHEP05(2012)120}{{\em JHEP} {\bf 1205} (2012)
   120},
\href{http://arxiv.org/abs/1109.0544}{{\tt arXiv:1109.0544 [hep-th]}}.
%%CITATION = ARXIV:1109.0544;%%.

\bibitem{DHH}
S.~Detournay, T.~Hartman, and D.~M. Hofman, ``{Warped Conformal Field
  Theory},'' \href{http://dx.doi.org/10.1103/PhysRevD.86.124018}{{\em
  Phys.Rev.} {\bf D86} (2012)  124018},
\href{http://arxiv.org/abs/1210.0539}{{\tt arXiv:1210.0539 [hep-th]}}.
%%CITATION = ARXIV:1210.0539;%%.

\bibitem{Hofman2}
D.~M. Hofman and B.~Rollier, ``{Warped Conformal Field Theory as Lower Spin
  Gravity},''
\href{http://arxiv.org/abs/1411.0672}{{\tt arXiv:1411.0672 [hep-th]}}.
%%CITATION = ARXIV:1411.0672;%%.

\bibitem{Clement}
K. Ait Moussa, G.~Cl\'ement, and C.~Leygnac, ``{The Black holes of topologically
  massive gravity},'' \href{http://dx.doi.org/10.1088/0264-9381/20/24/L01}{{\em
  Class.Quant.Grav.} {\bf 20} (2003)  L277--L283},
\href{http://arxiv.org/abs/gr-qc/0303042}{{\tt arXiv:gr-qc/0303042 [gr-qc]}}.
%%CITATION = GR-QC/0303042;%%.

\bibitem{dSCFT}
A.~Strominger, ``{The dS / CFT correspondence},''
  \href{http://dx.doi.org/10.1088/1126-6708/2001/10/034}{{\em JHEP} {\bf 0110}
  (2001)  034},
\href{http://arxiv.org/abs/hep-th/0106113}{{\tt arXiv:hep-th/0106113
  [hep-th]}}.
%%CITATION = HEP-TH/0106113;%%.

\bibitem{Minic}
V.~Balasubramanian, J.~de~Boer, and D.~Minic, ``{Mass, entropy and holography
  in asymptotically de Sitter spaces},''
  \href{http://dx.doi.org/10.1103/PhysRevD.65.123508}{{\em Phys.Rev.} {\bf D65}
  (2002)  123508},
\href{http://arxiv.org/abs/hep-th/0110108}{{\tt arXiv:hep-th/0110108
  [hep-th]}}.
%%CITATION = HEP-TH/0110108;%%.

\bibitem{LD}
G.~Compère, L.~Donnay, P.-H. Lambert, and W.~Schulgin, ``{Liouville theory
  beyond the cosmological horizon},''
  \href{http://dx.doi.org/10.1007/JHEP03(2015)158}{{\em JHEP} {\bf 1503} (2015)
   158},
\href{http://arxiv.org/abs/1411.7873}{{\tt arXiv:1411.7873 [hep-th]}}.
%%CITATION = ARXIV:1411.7873;%%.

\bibitem{NMG}
E.~A. Bergshoeff, O.~Hohm, and P.~K. Townsend, ``{Massive Gravity in Three
  Dimensions},'' \href{http://dx.doi.org/10.1103/PhysRevLett.102.201301}{{\em
  Phys.Rev.Lett.} {\bf 102} (2009)  201301},
\href{http://arxiv.org/abs/0901.1766}{{\tt arXiv:0901.1766 [hep-th]}}.
%%CITATION = ARXIV:0901.1766;%%.

\bibitem{Sandinista}
I.~Bengtsson and P.~Sandin, ``{Anti de Sitter space, squashed and stretched},''
  \href{http://dx.doi.org/10.1088/0264-9381/23/3/022}{{\em Class.Quant.Grav.}
  {\bf 23} (2006)  971--986},
\href{http://arxiv.org/abs/gr-qc/0509076}{{\tt arXiv:gr-qc/0509076 [gr-qc]}}.
%%CITATION = GR-QC/0509076;%%.


\bibitem{reboucas}
M.~Reboucas and J.~Tiomno, ``{On the Homogeneity of Riemannian Space-Times of
  Godel Type},''
\href{http://dx.doi.org/10.1103/PhysRevD.28.1251}{{\em Phys.Rev.} {\bf D28}
  (1983)  1251--1264}.
%%CITATION = PHRVA,D28,1251;%%.

%\cite{Rooman:1998xf}
\bibitem{Spindel} 
  M.~Rooman and P.~Spindel,
  ``Godel metric as a squashed anti-de Sitter geometry,''
  Class.\ Quant.\ Grav.\  {\bf 15}, 3241 (1998)
  [gr-qc/9804027].
  %%CITATION = GR-QC/9804027;%%
  %62 citations counted in INSPIRE as of 24 Apr 2015


\bibitem{Godel}
K.~G\"odel, ``{An Example of a new type of cosmological solutions of Einstein's
  field equations of graviation},''
\href{http://dx.doi.org/10.1103/RevModPhys.21.447}{{\em Rev.Mod.Phys.} {\bf 21}
  (1949)  447--450}.
%%CITATION = RMPHA,21,447;%%.

\bibitem{haw-ellis}
S.~Hawking and G.~Ellis,
``{The large scale structure of space-time},'' Cambridge University Press (1973).
%%CITATION = INSPIRE-87997;%%.

\bibitem{inta} 
  D.~Israel,
  ``Quantization of heterotic strings in a Godel / anti-de Sitter space-time and chronology protection,''
  JHEP {\bf 0401}, 042 (2004)
  [hep-th/0310158].
  %%CITATION = HEP-TH/0310158;%%
  %50 citations counted in INSPIRE as of 24 Apr 2015

\bibitem{instring}
S.~Detournay, D.~Orlando, P.~M. Petropoulos, and P.~Spindel,
  ``{Three-dimensional black holes from deformed anti-de Sitter},''
  \href{http://dx.doi.org/10.1088/1126-6708/2005/07/072}{{\em JHEP} {\bf 0507}
  (2005)  072},
\href{http://arxiv.org/abs/hep-th/0504231}{{\tt arXiv:hep-th/0504231
  [hep-th]}}.
%%CITATION = HEP-TH/0504231;%%.

\bibitem{intopologicallygaugetheories}
K. Ait Moussa and G.~Cl\'ement, ``{Topologically massive gravitoelectrodynamics:
  Exact solutions},'' \href{http://dx.doi.org/10.1088/0264-9381/13/8/023}{{\em
  Class.Quant.Grav.} {\bf 13} (1996)  2319--2328},
\href{http://arxiv.org/abs/gr-qc/9602034}{{\tt arXiv:gr-qc/9602034 [gr-qc]}}.
%%CITATION = GR-QC/9602034;%%.

\bibitem{israel}
  D.~Israel, C.~Kounnas, D.~Orlando and P.~M.~Petropoulos,
  ``Electric/magnetic deformations of S**3 and AdS(3), and geometric cosets,''
  Fortsch.\ Phys.\  {\bf 53}, 73 (2005)
  [hep-th/0405213].
  %%CITATION = HEP-TH/0405213;%%
  %62 citations counted in INSPIRE as of 24 Apr 2015


\bibitem{yotromas}
A.~Bouchareb and G.~Cl\'ement, ``{Black hole mass and angular momentum in
  topologically massive gravity},''
  \href{http://dx.doi.org/10.1088/0264-9381/24/22/018}{{\em Class.Quant.Grav.}
  {\bf 24} (2007)  5581--5594},
\href{http://arxiv.org/abs/0706.0263}{{\tt arXiv:0706.0263 [gr-qc]}}.
%%CITATION = ARXIV:0706.0263;%%.

\bibitem{3dorigin}
M.~Ba\~nados, G.~Barnich, G.~Comp\`ere, and A.~Gomberoff, ``{Three dimensional
  origin of Godel spacetimes and black holes},''
  \href{http://dx.doi.org/10.1103/PhysRevD.73.044006}{{\em Phys.Rev.} {\bf D73}
  (2006)  044006},
\href{http://arxiv.org/abs/hep-th/0512105}{{\tt arXiv:hep-th/0512105
  [hep-th]}}.
%%CITATION = HEP-TH/0512105;%%.

\bibitem{Clem}
G.~Cl\'ement, ``{Warped AdS(3) black holes in new massive gravity},''
  \href{http://dx.doi.org/10.1088/0264-9381/26/10/105015}{{\em
  Class.Quant.Grav.} {\bf 26} (2009)  105015},
\href{http://arxiv.org/abs/0902.4634}{{\tt arXiv:0902.4634 [hep-th]}}.
%%CITATION = ARXIV:0902.4634;%%.

\bibitem{inbigravity}
A.~F. Goya, ``{Anisotropic Scale Invariant Spacetimes and Black Holes in
  Zwei-Dreibein Gravity},''
  \href{http://dx.doi.org/10.1007/JHEP09(2014)132}{{\em JHEP} {\bf 1409} (2014)
   132},
\href{http://arxiv.org/abs/1406.4771}{{\tt arXiv:1406.4771 [hep-th]}}.
%%CITATION = ARXIV:1406.4771;%%.

\bibitem{inEinstein}
D.~Anninos, ``{Hopfing and Puffing Warped Anti-de Sitter Space},''
  \href{http://dx.doi.org/10.1088/1126-6708/2009/09/075}{{\em JHEP} {\bf 0909}
  (2009)  075},
\href{http://arxiv.org/abs/0809.2433}{{\tt arXiv:0809.2433 [hep-th]}}.
%%CITATION = ARXIV:0809.2433;%%.



\bibitem{Hohm-Tonni}
O.~Hohm and E.~Tonni, ``{A boundary stress tensor for higher-derivative gravity
  in AdS and Lifshitz backgrounds},''
  \href{http://dx.doi.org/10.1007/JHEP04(2010)093}{{\em JHEP} {\bf 1004} (2010)
   093},
\href{http://arxiv.org/abs/1001.3598}{{\tt arXiv:1001.3598 [hep-th]}}.
%%CITATION = ARXIV:1001.3598;%%.



%\cite{AyonBeato:2009yq}
\bibitem{Sch}
  E.~Ayon-Beato, G.~Giribet and M.~Hassaine,
  ``Bending AdS Waves with New Massive Gravity,''
  JHEP {\bf 0905}, 029 (2009)
  [arXiv:0904.0668 [hep-th]].
  %%CITATION = ARXIV:0904.0668;%%
  %72 citations counted in INSPIRE as of 24 Apr 2015
%\cite{AyonBeato:2009nh}

\bibitem{Lif}
  E.~Ayon-Beato, A.~Garbarz, G.~Giribet and M.~Hassaine,
  ``Lifshitz Black Hole in Three Dimensions,''
  Phys.\ Rev.\ D {\bf 80}, 104029 (2009)
  [arXiv:0909.1347 [hep-th]]
  %%CITATION = ARXIV:0909.1347;%%
  %104 citations counted in INSPIRE as of 24 Apr 2015

%\cite{Clement:2009ka}
\bibitem{Clemooo} 
  G.~Clement,
  ``Black holes with a null Killing vector in new massive gravity in three dimensions,''
  Class.\ Quant.\ Grav.\  {\bf 26}, 165002 (2009)
  [arXiv:0905.0553 [hep-th]].
  %%CITATION = ARXIV:0905.0553;%%
  %42 citations counted in INSPIRE as of 24 Apr 2015

\bibitem{Troncoso}
J.~Oliva, D.~Tempo, and R.~Troncoso, ``{Three-dimensional black holes,
  gravitational solitons, kinks and wormholes for BHT massive gravity},''
  \href{http://dx.doi.org/10.1088/1126-6708/2009/07/011}{{\em JHEP} {\bf 0907}
  (2009)  011},
\href{http://arxiv.org/abs/0905.1545}{{\tt arXiv:0905.1545 [hep-th]}}.
%%CITATION = ARXIV:0905.1545;%%.


%\cite{Siampos:2013foa}
\bibitem{Siampos} 
  K.~Siampos and P.~Spindel,
  ``Solutions of massive gravity theories in constant scalar invariant geometries,''
  Class.\ Quant.\ Grav.\  {\bf 30}, 145014 (2013)
  [arXiv:1302.6250 [hep-th]].
  %%CITATION = ARXIV:1302.6250;%%
  %2 citations counted in INSPIRE as of 13 May 2015


\bibitem{BY}
J.~D. Brown and J.~York, James~W., ``{Quasilocal energy and conserved charges
  derived from the gravitational action},''
  \href{http://dx.doi.org/10.1103/PhysRevD.47.1407}{{\em Phys.Rev.} {\bf D47}
  (1993)  1407--1419},
\href{http://arxiv.org/abs/gr-qc/9209012}{{\tt arXiv:gr-qc/9209012 [gr-qc]}}.
%%CITATION = GR-QC/9209012;%%.

\bibitem{GiribetGoya}
G.~Giribet and A.~Goya, ``{The Brown-York mass of black holes in Warped Anti-de
  Sitter space},'' \href{http://dx.doi.org/10.1007/JHEP03(2013)130}{{\em JHEP}
  {\bf 1303} (2013)  130},
\href{http://arxiv.org/abs/1212.2100}{{\tt arXiv:1212.2100 [hep-th]}}.
%%CITATION = ARXIV:1212.2100;%%.

\bibitem{NMG2}
E.~A. Bergshoeff, O.~Hohm, and P.~K. Townsend, ``{More on Massive 3D
  Gravity},'' \href{http://dx.doi.org/10.1103/PhysRevD.79.124042}{{\em
  Phys.Rev.} {\bf D79} (2009)  124042},
\href{http://arxiv.org/abs/0905.1259}{{\tt arXiv:0905.1259 [hep-th]}}.
%%CITATION = ARXIV:0905.1259;%%.

\bibitem{Barnich:2001jy}
G.~Barnich and F.~Brandt, ``{Covariant theory of asymptotic symmetries,
  conservation laws and central charges},''
  \href{http://dx.doi.org/10.1016/S0550-3213(02)00251-1}{{\em Nucl.Phys.} {\bf
  B633} (2002)  3--82},
\href{http://arxiv.org/abs/hep-th/0111246}{{\tt arXiv:hep-th/0111246
  [hep-th]}}.
%%CITATION = HEP-TH/0111246;%%.

\bibitem{Barnich:2007bf}
G.~Barnich and G.~Comp\`ere, ``{Surface charge algebra in gauge theories and
  thermodynamic integrability},''
  \href{http://dx.doi.org/10.1063/1.2889721}{{\em J.Math.Phys.} {\bf 49} (2008)
   042901},
\href{http://arxiv.org/abs/0708.2378}{{\tt arXiv:0708.2378 [gr-qc]}}.
%%CITATION = ARXIV:0708.2378;%%.

\bibitem{Nam}
S.~Nam, J.-D. Park, and S.-H. Yi, ``{Mass and Angular momentum of Black Holes
  in New Massive Gravity},''
  \href{http://dx.doi.org/10.1103/PhysRevD.82.124049}{{\em Phys.Rev.} {\bf D82}
  (2010)  124049},
\href{http://arxiv.org/abs/1009.1962}{{\tt arXiv:1009.1962 [hep-th]}}.
%%CITATION = ARXIV:1009.1962;%%.


%\cite{Donnay:2015iia}
\bibitem{LG} 
  L.~Donnay and G.~Giribet,
  ``Holographic entropy of Warped-AdS$_3$ black holes,'' \href{http://dx.doi.org/10.1007/JHEP06(2015)099}{{\em JHEP}
  {\bf 1506} (2015)  099},
  \href{http://arxiv.org/abs/1504.05640}{{\tt arXiv:1504.05640 [hep-th]}}.
  %%CITATION = ARXIV:1504.05640;%%


\end{thebibliography}
\end{document}